\documentclass{article}
\usepackage{amssymb}
\usepackage{amsmath}
\usepackage{amsfonts}

\input{tcilatex}
\begin{document}

\title{Semiclassical analysis of defect sine-Gordon theory\\
\medskip }
\author{F. Nemes\medskip \\
Institute for Theoretical Physics\\
Roland E\"{o}tv\"{o}s University,\\
H-1117 P\'{a}zm\'{a}ny s. 1/A, Budapest, Hungary\medskip }
\maketitle

\begin{abstract}
The classical sine-Gordon model is a two-dimensional integrable field
theory, with particle like solutions the so-called solitons. Using its
integrability one can define its quantum version without the process of
canonical quantization. This bootstrap method uses the fundamental
propterties of the model and its quantum features in order to restrict the
structure of the scattering matrix as far as possible. The classical model
can be extended with integrable discontinuities, purely transmitting
jump-defects. Then the quantum version of the extended model can be
determined via the bootstrap method again. But the outcoming quantum theory
contains the so-called CDD uncertainity. The aim of this article is to carry
throw the semiclassical approximation in both the classical and the quantum
side of the defect sine-Gordon theory. The CDD ambiguity can be restricted
by comparing the two results. The relation between the classical and quantum
parameters as well as the resoncances appeared in the spectrum are other
objectives. \pagebreak
\end{abstract}

\section{Introduction}

Recently, there has been an increasing interest in\ integrable quantum field
theories including defects or impurities. The motivations come from both the
theoretical and the application part of physics. In the solid state and in
the statistical physics there are realistic physical applications.\ There is
also a need of theoretical understanding of this kind of field theory, which
gives a well defined model for impurities.

Due to the no-go theorem formulated by Delfino, Mussardo and Simonetti the
defect theories stay in the background for some time \cite{bs1}, \cite{bs2}.
The theorem states that a relativistically invariant theory with an
integrable non-free interaction in the bulk region permits only two types of
defects: the purely transmitting and the purely reflecting ones. This
theorem was originally formulated for diagonal theories, and later it was
extended to many type of non-diagonal ones \cite{bs3}. (Although some effort
has been made to overcome this obstacle by giving up the Lorentz invariance,
see for instance \cite{bs4} and references therein).

As the no-go theorem showed non-free integrable defect theories are purely
transmitting. This proof retains the researchers for some time to analyze
these models. But the subject were revived after founding explicit
Lagrangian realizations for the defect theories \cite{bs17}. Following the
original idea many integrable defect theories were constructed at the
classical level \cite{bs18}, \cite{bs19}, \cite{bs20}. The basis for the
quantum formulation of defect theories is provided by the folding trick \cite%
{bs21} by which one can map any defect theory into a boundary one. As a
consequence defect unitarity, defect crossing symmetry and defect bootstrap
equations together with defect Coleman-Thun mechanism are derived. Despite
of these results the explicitly solved relativistically invariant defect
quantum field theories are quite rare, containing basically the sine-Gordon
and affine Toda field theories \cite{bs22}, \cite{bs23}, \cite{bs24}. The
explicit solution of the Lee-Yang and sinh-Gordon theories are also known 
\cite{bs25}.

It is required that the transmission matrix of the defect theory satisfy the
Yang-Baxter equation (also known as "factorisability condition"), and the
standard equation of unitarity and crossing symmetry. These equations have
enough restrictive power to determine the T-matrix up to the so-called "CDD
ambiguity".

It means that there is an unknown scalar function in the scattering matrix
and - a different one - in the defect transmission matrix. These functions
cannot be fixed via the bootstrap method. One way to restrict this ambiguity
is to calculate the phase shift by the semiclassical approximation. In the
case of quantum sine-Gordon theory the classical phase shift has been
compared with the scattering matrix earlier \cite{key-15}, \cite{key-24}.
One of the aims of this article is to determine the phase shift in the case
of \textit{defect} sine-Gordon theory. Using the result we can anchor the
CDD uncertainty. The defect theory has new symmetries, so we have to
recalculate the time delay for the new cases.

Another objective is to map the pole structure of the T-matrix. In this way
we can study the spectrum of the theory. This direct examination can show if
there are stable or unstable bound states or resonances in the model.

This paper is organized as follows: In section 2, we explain the classical
defect sine-Gordon theory, where we give the time delay for the case of odd
topological charge. In section 3 we introduce the defect quantum sine-Gordon
theory and outline the semiclassical approximation. In section 4 we perform
the semiclassical approximation on the basis of the previous sections.
Firstly we start from the classical side. Then we calculate the phase shift
from the transmission matrix in the case of the even and odd charged
defects. We analyze the matrix structure in the same limit which lead us to
analyze the spectrum directly in section 5. Finally we conclude in section 6
and give directions for further research.

\section{Classical defect sine-Gordon theory}

In the bulk the sine-Gordon model can be defined by the following Lagrangian
density:%
\begin{equation}
\mathcal{L}_{_{\Phi }}:=\frac{1}{2}\left( \left( \partial _{t}\Phi \right)
^{2}-\left( \partial _{x}\Phi \right) ^{2}\right) +\frac{m^{2}}{\beta ^{2}}%
(\cos (\beta \Phi )-1)  \label{lagrangian for the bulk}
\end{equation}

The integrability follows from the continuity equations for an infinite set
of local currents. For classical considerations it is often convenient to
remove the mass parameter $m$ and the coupling $\beta $ by rescaling the
field. The Lagrangian is invariant under the well known bulk symmetries%
\begin{equation}
\Phi \rightarrow \Phi +\frac{2n\pi }{\beta },\;n\in \mathbb{Z}
\end{equation}%
, which interpolate among the different ground states of the bulk.

A single jump-defect placed at $x=0$ is described by modifying the
Lagrangian. Firstly we define two fields: the $u$ field for the $x<0$ and $v$
for the $x>0$ part of the $x$-axis. The "time evolution" of these two
restricted fields will follow the bulk region. The defect works as an
internal boundary linking the two fields: a delta function contribution at $%
x=0$ couples $u$ and $v$. The detailed Lagrangian density is the following 
\cite{nf1}%
\begin{equation}
\mathcal{L}:=\Theta (-x)\mathcal{L}_{u}+\Theta (x)\mathcal{L}_{v}-\delta (x)%
\left[ \frac{1}{2}\left( u\partial _{t}v-v\partial _{t}u\right) -\mathcal{B}%
(u,v)\right]  \label{lagrangian for the defect}
\end{equation}

where the defect potential is%
\begin{equation}
\mathcal{B}=-\frac{2m\sigma }{\beta ^{2}}\cos \left( \beta \frac{u+v}{2}%
\right) -\frac{2m}{\sigma \beta ^{2}}\cos \left( \beta \frac{u-v}{2}\right)
\label{defect potential}
\end{equation}

It follows that the bulk field equations - after the suitable rescaling -
for $u$ and $v$ are:%
\begin{equation}
\begin{array}{c}
x<0:\;\partial _{\mu }\partial ^{\mu }u=-\sin (u) \\ 
x>0:\;\partial _{\mu }\partial ^{\mu }v=-\sin (v)%
\end{array}
\label{equation of motion}
\end{equation}

For the $x=0$ point we get the defect conditions:%
\begin{equation}
\begin{array}{c}
\partial _{x}u-\partial _{t}v=-\sigma \sin \left( \frac{u+v}{2}\right) -%
\frac{1}{\sigma }\sin \left( \frac{u-v}{2}\right) \\ 
\partial _{x}v-\partial _{t}u=\sigma \sin \left( \frac{u+v}{2}\right) -\frac{%
1}{\sigma }\sin \left( \frac{u-v}{2}\right)%
\end{array}
\label{defect conditions}
\end{equation}

The first order time derivatives in the defect Lagrangian is required by
integrability. The theory has an infinite set of mutually commuting
integrals of motion.

The period of the defect potential is twice of the bulk potential. So the
new Lagrangian is not invariant under the bulk symmetries $u\rightarrow
u+2a\pi /\beta ,$ $v\rightarrow v+2b\pi /\beta $ where $a$ and $b$ are
integers, The new symmetries are%
\begin{equation}
u\rightarrow u+4a\pi /\beta ,v\rightarrow v+4b\pi /\beta ,\;a,b\in \mathbb{Z}
\label{new symmetries}
\end{equation}

It is not even invariant under the reflections $u\rightarrow -u$ or $%
v\rightarrow -v$. But $\left( \ref{lagrangian for the defect}\right) $ is
invariant under certain combinations of the earlier transformations, such as
reflecting both fields simultaneously.

The Lagrangian does not violate time translation. It follows that the total
energy $\mathcal{E}$ will be conserved. In a certain time this constant $%
\mathcal{E}$ consists of the energy of both fields, and a contribution from
the defect. This contribution is a function of the fields evaluated at the
place of the defect%
\begin{equation}
\mathcal{E}=E\left( u\right) +E\left( v\right) +B\left( u,v\right)
\end{equation}%
A single soliton of the sine-Gordon model may be described by 
\begin{equation}
u(t,x):=\pm \frac{4}{\beta }\arctan \left( e^{-m\frac{x+vt}{\sqrt{1-v^{2}}}%
}\right) \,  \label{soliton solution trigonometric form}
\end{equation}%
where $+$ stands for solitons, and $-$ for anti-solitons, and the $v$
quantity is the velocity of the soliton. The energy of this soliton from the
Hamilton density 
\begin{equation}
E=\frac{M}{\sqrt{1-v^{2}}}=M\cosh (\theta )  \label{energy of one soliton}
\end{equation}%
where $\theta $ is the rapidity, $M=8m/\beta ^{2}$.

It is useful to write the above solution in the following - equivalent -
form \cite{nf1}%
\begin{equation}
e^{iu/2}=\frac{1\pm iE}{1\mp iE},\;E=e^{ax+bt+c}
\label{soliton solution exponential form for u}
\end{equation}%
where in terms of rapidity $a=\cosh (\theta ),b=-\sinh (\theta )$ are real
quantities and clearly $a^{2}-b^{2}=1$. Now the upper sign stands for
solitons. Notice that there is no reason why $u=v$ at $x=0$, that is the
whole field is not necessarily continuous.

Suppose there is a soliton moving in a positive sense along the $x$-axis
described by the $u$ field. It encounters the defect, then a similar, but
delayed, $v$ field soliton emerges. We suppose that the solution for the $v$
field 
\begin{equation}
e^{iv/2}=\frac{1+ziE}{1-ziE},\;E=e^{ax+bt+c}
\label{soliton solution exponential form for v}
\end{equation}%
where $z$ contains the time delay in this manner 
\begin{equation}
zE=e^{ax+bt+c+ln(z)}\Rightarrow \Delta t=\frac{ln(z)}{b}
\label{zbol az idokeses}
\end{equation}

If we rewrite the defect parameter $\sigma =e^{-\eta }$, the expression for $%
z$ is the following%
\begin{equation}
z=\frac{e^{-\theta }+\sigma }{e^{-\theta }-\sigma }=\coth \left( \frac{\eta
-\theta }{2}\right)
\end{equation}%
which can be substituted into $\left( \ref{zbol az idokeses}\right) $. The
time delay, which was calculated already in \cite{nf1} 
\begin{equation}
\Delta t=\frac{\ln \left( \tanh \left( \frac{\eta -\theta }{2}\right)
\right) }{\sinh \left( \theta \right) }
\label{time delay with the presence of a defect}
\end{equation}

Suppose that $\theta >0$ and $\eta <0$. In this case $z<0$, which means that
the incoming soliton flips to an anti-soliton. But in the $\eta >0$ case
there are several possibilities. If $\theta <\eta $ the type of soliton
remains unchanged; if $\theta >\eta $ the soliton type changes on the
defect. In the special $\theta =\eta $ case the incoming soliton is delayed
infinitely, in effect the incoming soliton is bounded. In this situation the
incoming soliton can be replaced in the far future by the static $u=0$, $%
v=2\pi $ solution. This solution stores the same energy and momentum into
the defect as the incoming soliton has, so the defect works like a capacitor
in an electrical circuit.

We can generalize this idea, since the two static fields $u=2\pi n$ and $%
v=2\pi m$ satisfies the $\left( \ref{equation of motion}\right) $ equations
of motion and the $\left( \ref{defect conditions}\right) $ defect conditions
if $n$ and $m$ are integers. The conserved topological charge is%
\begin{equation*}
Q:=\frac{\beta }{2\pi }\dint\limits_{-\infty }^{\infty }dx\partial _{x}\phi =%
\frac{\beta }{2\pi }(\underset{x\rightarrow +\infty }{\lim }v(x)-\underset{%
x\rightarrow -\infty }{\lim }u(x)+\underset{x\rightarrow +0}{\lim }v(x)-%
\underset{x\rightarrow -0}{\lim }u(x))=n-m
\end{equation*}%
where the limit at the origin measures the strength of the defect \cite{bs17}%
.

We have to discuss the time delay in a new unnamed case : we have to take
care of the parity of the charge, because of the new $\left( \ref{new
symmetries}\right) $ symmetries. To show this let $v\rightarrow v+2\pi $,
while $u$ remains unchanged

\begin{equation}
e^{iv/2}=-\frac{1+ziE}{1-ziE}
\end{equation}

In this case the defect links solitons differing by one unit of charge.
Suppose the situation $ax+bt+c=0$ so $E=1$. Substituting this solution into
the defect conditions the equation for $z$ is 
\begin{multline}
\frac{4a}{2}-\frac{4bz}{1+z^{2}}=\frac{1}{2}i\sigma \left( \frac{\left(
1+i\right) \left( 1+iz\right) }{\left( 1-i\right) \left( 1-iz\right) }-\frac{%
\left( 1-i\right) \left( 1-iz\right) }{\left( 1+i\right) \left( 1+iz\right) }%
\right) + \\
\frac{1}{2}\frac{i}{\sigma }\left( \frac{\left( 1+i\right) \left(
1-iz\right) }{\left( 1-i\right) \left( 1+iz\right) }-\frac{\left( 1-i\right)
\left( 1+iz\right) }{\left( 1+i\right) \left( 1-iz\right) }\right)
\end{multline}

After some simplification we get%
\begin{equation}
z=\tanh \left( \frac{\eta -\theta }{2}\right)  \label{z on the defect}
\end{equation}

The corresponding time delay is negative of that of the even case%
\begin{equation}
\Delta t=-\frac{\ln (\tanh (\frac{\eta -\theta }{2}))}{\sinh (\theta )}
\label{time delay on the odd type defect}
\end{equation}

There are N-soliton type solutions in the bulk, which can be generated by
Hirota's method. In this article we use only the two-soliton solutions. A
soliton-antisoliton pair - moving with the relative velocity $v$ - can be
written%
\begin{equation}
\Phi _{s\overline{s}}=\frac{4}{\beta }\arctan \left( \frac{\sinh (m\gamma vt)%
}{v\cosh (m\gamma x)}\right) ,\;\gamma =1/\sqrt{1-v^{2}}
\label{soliton antisoliton solution}
\end{equation}

From this expression we can easily obtain the well known time delay 
\begin{equation}
\Delta t=\frac{2ln(v)}{m\gamma v}=\frac{2\ln (\tanh (\theta ))}{m\sinh
(\theta )},\;\theta =\theta _{2}-\theta _{1}
\label{time delay in velocity and rapidity}
\end{equation}%
with the relative $\theta $ rapidity. The presence of the time delay shows
that there is an interaction between solitons.

Worthy of note that the $\left( \ref{time delay in velocity and rapidity}%
\right) $ time delay is exactly twice of the $\left( \ref{time delay with
the presence of a defect}\right) $ time delay on the defect (if formally we
think of the $\eta $ defect parameter as one of the rapidity).

\section{Quantum defect sine-Gordon theory}

In the quantum theory of the sine-Gordon model the $\mathcal{H}$
Hilbert-space is the Fock-space of multiparticle states. The vectors of $%
\mathcal{H}$ is generated "particle creation operators" $A_{a}\left( \theta
\right) $%
\begin{equation}
\left\vert A_{a_{1}}\left( \theta _{1}\right) A_{a_{2}}\left( \theta
_{2}\right) ...A_{a_{n}}\left( \theta _{n}\right) \right\rangle
=A_{a_{1}}\left( \theta _{1}\right) A_{a_{2}}\left( \theta _{2}\right)
...A_{a_{n}}\left( \theta _{n}\right) \left\vert 0\right\rangle
\label{an arbitrary state vector}
\end{equation}

These vectors can be interpreted as the asymptotic ("in-" or "out-")
scattering states \cite{bs5}, which correspond to the limit of the soliton
type solutions in the far past and in the far future. In the defect theory
we can describe the purely transmitting defect with a matrix which relates
these "in-" and "out-" vectors of $\mathcal{H}$.

We could see using $\left( \ref{time delay with the presence of a defect}%
\right) $ that a classical defect can delay a soliton or an anti-soliton
infinitely and the final state might be replaced by a static one. In the
corresponding quantum theory we have to say the defect "bounds" this
particle. Reversing the sense of time it appears that defects can decay and
might produce particles resemblance to an excited atom. It is clear that an
incoming soliton can induce a decay.

We have to take into account also that in the defect theory the bulk
symmetries are broken. This idea lead us to distinguish the even and odd
charged defects. In the corresponding quantum theory we have to introduce
different transfer matrices for them.

Finally, the notation for the even type will be \cite{nf1}%
\begin{equation}
^{e}T_{a\alpha }^{b\beta }(\theta )
\label{the notation of the transmission matrix}
\end{equation}%
where $a$ labels the incoming particle and $b$ the outgoing one. Its value
may be $+$ for a soliton or $-$ for an anti-soliton. The $\alpha $ and $%
\beta $ numbers are even integer; they sign the topological charge of the in
and out particles and $\beta -\alpha $ is the even defect charge. This
matrix regarded as describing the transmission of a particle with rapidity $%
\theta $\ from the $x<0$ region to the region $x>0$. The structure of this
matrix can be explored via the bootstrap method.

Firstly, these families of even transmission matrices should satisfy the
unitarity condition (for real rapidity)%
\begin{equation}
^{e}T(\theta )^{e}T^{\dagger }(\theta )=1  \label{unitarity condition}
\end{equation}

The transmission matrices relate states of the system in the far future to
those in the far past and the state of the system is labelled by the soliton
rapidity and its topological charge together.

We have to invoke the well known sine-Gordon S-matrix for a two-body
scattering process in the bulk \cite{key-21}. It is given by%
\begin{equation}
S_{kl}^{mn}(\theta ):=\varrho (\theta )\left( 
\begin{array}{cccc}
a(\theta ) & 0 & 0 & 0 \\ 
0 & c(\theta ) & b(\theta ) & 0 \\ 
0 & b(\theta ) & c(\theta ) & 0 \\ 
0 & 0 & 0 & a(\theta )%
\end{array}%
\right)
\end{equation}%
where $k,l$ label the incoming particles and $m,n$ the outgoing ones. $%
\theta =\theta _{1}-\theta _{2}$ where the $\theta _{1}$ rapidity belong to
the $k,n$ and $\theta _{2}$ is the rapidity of the $l,m$ particles. The
various pieces of the matrix are defined by: 
\begin{equation}
a(\theta ):=e^{i\pi \gamma }e^{-\gamma \theta }-e^{-i\pi \gamma }e^{\gamma
\theta },\;b(\theta ):=e^{\gamma \theta }-e^{-\gamma \theta },\;c(\theta
):=e^{i\pi \gamma }-e^{-i\pi \gamma }  \label{eq:atszorzok}
\end{equation}

In this notation the crossing property of the S-matrix is represented by

\begin{equation}
S_{kl}^{mn}(i\pi -\theta )=S_{k~-m}^{-l~n}(\theta )
\end{equation}

The overall $\varrho \left( \theta \right) $ function is 
\begin{equation}
\varrho (\theta ):=\frac{\Gamma (1+i\gamma \theta /\pi )\Gamma (1-\gamma
-i\gamma \theta /\pi )}{2\pi i}\prod_{k=1}^{\infty }R_{k}(\theta )R_{k}(i\pi
-\theta )
\end{equation}%
where%
\begin{equation}
R_{k}(\theta )=\frac{\Gamma (2k\gamma +i\gamma \theta /\pi )\Gamma
(1+2k\gamma +i\gamma \theta /\pi )}{\Gamma ((2k+1)\gamma +i\gamma \theta
/\pi )\Gamma (1+(2k-1)\gamma +i\gamma \theta /\pi )}  \label{eq:rkteta}
\end{equation}%
The conventions adopted by Konik and LeClair \cite{bs22} have been used.
Therefore the coupling $\gamma $ in terms of the Lagrangian coupling $\beta $
is given by%
\begin{equation}
\frac{1}{\gamma }=\frac{\beta ^{2}}{8\pi -\beta ^{2}}
\label{coupling constants}
\end{equation}

Since the defect is purely transmitting, the heuristic arguments based on
factorisability and bulk integrability would require \cite{bs1}%
\begin{equation}
S_{kl}^{mn}(\theta )^{e}T_{n\alpha }^{t\beta }(\theta _{1})^{e}T_{m\beta
}^{s\gamma }(\theta _{2})=\text{ }^{e}T_{l\alpha }^{n\beta }(\theta
_{2})^{e}T_{k\beta }^{m\gamma }(\theta _{1})S_{mn}^{st}(\theta )
\label{YB condition}
\end{equation}

It is useful to structure the transmission matrix into block matrices. The
entries are infinite dimensional block matrices labelled by the topological
charge of the defect%
\begin{equation}
^{e}T=\left( 
\begin{array}{cc}
T_{+}^{+} & T_{+}^{-} \\ 
T_{-}^{+} & T_{-}^{-}%
\end{array}%
\right) \equiv \left( 
\begin{array}{cc}
A & B \\ 
C & D%
\end{array}%
\right)  \label{even matrix basic structure}
\end{equation}

Using the triangle relation $\left( \ref{YB condition}\right) $ and a
collection of general principles they following result can be obtained \cite%
{nf1}%
\begin{equation}
\left( 
\begin{array}{cc}
A_{\alpha }^{\beta } & B_{\alpha }^{\beta } \\ 
C_{\alpha }^{\beta } & D_{\alpha }^{\beta }%
\end{array}%
\right) :=f(q,x)\left( 
\begin{array}{cc}
\frac{1}{\sqrt{\nu }}Q^{\alpha }\delta _{\alpha }^{\beta } & e^{-i\pi \gamma
/2+\gamma (\theta -\eta )}\delta _{\alpha }^{\beta -2} \\ 
e^{-i\pi \gamma /2+\gamma (\theta -\eta )}\delta _{\alpha }^{\beta +2} & 
\sqrt{\nu }Q^{-\alpha }\delta _{\alpha }^{\beta }%
\end{array}%
\right)  \label{T matrix even}
\end{equation}

The definition of the overall $f$ function is given by%
\begin{equation}
f(q,x):=\frac{e^{i\pi (1+\gamma )/4}}{(1+ipx)}\frac{r(\theta )}{r(-\theta )}%
,\quad px:=e^{\gamma (\theta -\eta )}  \label{common  function}
\end{equation}

where 
\begin{equation}
r(x)=\prod_{k=0}^{\infty }\frac{\Gamma (k\gamma +1/4-i\gamma \tilde{\theta}%
/2\pi )\Gamma ((k+1)\gamma +3/4-i\gamma \tilde{\theta}/2\pi )}{\Gamma
((k+1/2)\gamma +1/4-i\gamma \tilde{\theta}/2\pi )\Gamma ((k+1/2)\gamma
+3/4-i\gamma \tilde{\theta}/2\pi )},\tilde{\theta}:=\theta -\eta
\label{gammas product in the even T matrix}
\end{equation}

The examination of the poles of $^{e}T$ indicates that some pole corresponds
to an unstable defect bound state. For this reason, there will be a
bootstrap condition linking the the even and odd transmission matrices. In
detail%
\begin{equation}
c_{b\alpha }^{\gamma }{}^{o}T_{a\gamma }^{c\delta }=S_{ab}^{pq}\left( \theta
-\eta +\frac{i\pi }{2\gamma }\right) {}^{e}T_{q\alpha }^{c\beta }c_{p\beta
}^{\delta }  \label{the conection between even T and odd T}
\end{equation}

For the odd case we use the notation $\hat{A}$ in place of $A$. From the
bootstrap condition we get%
\begin{equation}
\hat{A}(\theta )=\frac{1}{\sqrt{\nu }}\frac{e^{-i\pi (1+\gamma )/4}}{1+ipx}%
\frac{\cos (\pi /4\gamma -i(\theta -\eta )/2)}{\sin (\pi /4\gamma -i(\theta
-\eta )/2)}\frac{s(x)}{\overline{s}(x)}  \label{odd T matrix elem for A}
\end{equation}

where $s(x)$ is very similar to $r(x)$%
\begin{equation}
s(x)=\prod\limits_{k=0}^{\infty }\frac{\Gamma (k\gamma +3/4+i\gamma \tilde{%
\theta}/2\pi )\Gamma ((k+1)\gamma +1/4+i\gamma \tilde{\theta}/2\pi )}{\Gamma
((k+1/2)\gamma +1/4+i\gamma \tilde{\theta}/2\pi )\Gamma ((k+1/2)\gamma
+3/4+i\gamma \tilde{\theta}/2\pi )}  \label{common function in the odd case}
\end{equation}

It is clear that the \textit{matrix equation}s above do not have a unique
solution. One can multiply a given solution with a proper scalar function
factor and the result remains a solution. This scalar function must satisfy
the unitarity condition and the crossing property 
\begin{eqnarray*}
f(\theta )f(-\theta ) &=&1 \\
f(i\pi -\theta ) &=&f(\theta )
\end{eqnarray*}

This undetermined scalar function is the referred "CDD\ ambiguity".

\section{Semiclassical comparison}

Suppose that $\Psi $ is the wave function of the \textit{in-state }which is
given in the $\left( \ref{an arbitrary state vector}\right) $ form. The
"defect" theory and the "bulk" theory yield different out-states because of
the interaction with the defect. Now we turn to calculate the shifted phase
in the "defect" theory from the classical and from the quantum model. Then
we compare the results to restrict the scalar function of the "CDD\
ambiguity" in question.

\subsection{Starting from the classical side}

To calculate the phase shift from the classical model we use the
semiclassical approximation. The semiclassical method is worked out for
particle quantum mechanics but it can be generalized for the field theory.
In this paper the method of Jackiw and Woo will be followed \cite{key-15}.

Let's write the Schr\"{o}dinger equation 
\begin{equation}
\hat{H}\Psi =E\Psi  \label{eq:schrodinger}
\end{equation}%
then make the usual assumption that the wave function has the following
separate form%
\begin{equation}
\Psi (x,t)=A(x,t)e^{\frac{i}{\hbar }S(x,t)},  \label{eq:szemiklasszikus}
\end{equation}

The most important guess: in space and time the amplitude $A(x,t)$ changes
much more slowly than the phase $S(x,t)/\hbar $. At first look it seems that
this corresponds to the limit $\hbar \rightarrow 0.$ But - following from
the series expansion - $\Psi $ has a relevant singularity at this point. So
it is better to say that $S\gg \hbar $, or - in field theory - the coupling
constant is supposed to be very small.

Using $\left( \ref{eq:schrodinger}\right) $ we get in the leading order%
\begin{equation}
S(x)=\int_{x_{0}}^{x_{f}}p(x,E)dx+\hbar \varphi _{0}  \label{hatas}
\end{equation}%
where $p(x,E)$ is the momentum affected by the presence of the potential.
The known result for the semiclassical phase shift%
\begin{equation}
2\delta \left( E\right) =\lim \int_{x_{0}}^{x_{f}}dx(p(x,E)-p(E))
\end{equation}

But starting from the Lagrangian one can derive a connection between the
time delay and the momentum of the \textit{classical} particle 
\begin{equation}
\Delta t\left( E\right) =\frac{\partial }{\partial E}%
\int_{x_{0}}^{x_{f}}dx(p(x,E)-p(E))  \label{idokeses}
\end{equation}%
where $p(E)$ is the momentum of the free particle. This is the time of
flight for a classical particle moving from position $x_{0}$ to $x_{f}$. In
the case of unbounded motion we have to take the limit $x_{0}\rightarrow
\infty ,x_{f}\rightarrow \infty .$

Comparing the last two formulas the important result is the following

\begin{equation*}
\delta (E)-\delta (E_{th})=\frac{1}{2}\int_{E_{th}}^{E}dE^{\prime }\triangle
t(E^{\prime })
\end{equation*}

Jackiw and Wo generalize this idea for the field theory: the formula remains
unchanged but the $x_{0}$ and $x_{f}$ is replaced by the $\Phi _{i}$ initial
and $\Phi _{f}$ final field configurations. The time of flight means the
time of field evolution between the $\Phi _{i}$ and $\Phi _{f}$
configurations in the presence of a potential. The phase shift formula for
the \textit{field} theory \cite{key-15}

\begin{equation}
\delta (E)=n_{B}\pi +\int_{E_{th}}^{E}dE^{\prime }\triangle t(E^{\prime })
\label{semiclassical phase shift from classical  time delay}
\end{equation}

In this formula the lower bound of the integral $E_{th}$ has a meaning : it
is the threshold energy where $p\left( E_{th}\right) =0$. Below the
threshold energy the time delay is infinite, so the field cannot reach the
final stage. Hence we are interested in the case when the integral starts 
\textit{just above} $E_{th}$. The bound states start somewhere just below $%
E_{th}$. Its phase shift contribution is taken into account with $n_{B}$ the
number of bound states.

\medskip Now we only have to substitute the time delay $\left( \ref{time
delay with the presence of a defect}\right) $ into the previous formula $%
\left( \ref{semiclassical phase shift from classical time delay}\right) $.
To do this one must change the variable of integration from the energy $E$
to the rapidity $\theta $, and we need an expression for%
\begin{equation}
\frac{\partial E\left( \theta \right) }{\partial \theta }
\label{the derivative of the energy}
\end{equation}

The $\left( \ref{energy of one soliton}\right) $ formula is the energy of
one soliton in the "bulk". In the defect case the energy is conserved, hence
the whole energy can be evaluated at every position of the soliton. Suppose
that the soliton is far from the defect. In this case the energy stored in
the defect is a constant value: independent from $\theta $. So in the $%
\left( \ref{the derivative of the energy}\right) $ derivative we can use the
formula for the "bulk".

The hyperbolic function in the denominator of the $\left( \ref{time delay
with the presence of a defect}\right) $ time delay disappears because%
\begin{equation}
\text{ }\frac{\partial E}{\partial \theta }=M\sinh (\theta )
\label{variable substitution below  the integral}
\end{equation}

After the substitution the result from the classical side%
\begin{equation}
\delta (E)=n_{B}\pi +\frac{8m}{\beta ^{2}}\dint\limits_{0}^{\theta }d\theta
^{\prime }\ln \left[ \tanh \frac{\left( \eta -\theta ^{\prime }\right) }{2}%
\right]  \label{phase shift on the even defect}
\end{equation}

\subsection{Starting from the quantum side}

Now we turn to the quantum field theory, where we know the even transmission
matrix $\left( \ref{even matrix basic structure}\right) $. The matrix
entries, especially the common $\left( \ref{common function}\right) $ $f$
function, are product of gamma functions. Accordingly to extract the phase
shift we use an integral representation of the natural logarithm of gamma
functions

\begin{equation}
\ln (\Gamma (\zeta ))=\int_{0}^{\infty }\frac{dt}{t}e^{-t}\left[ \zeta -1+%
\frac{e^{-(\zeta -1)t}-1}{1-e^{-t}}\right] ,\mathfrak{\;Re(\zeta )}>0
\label{integral representation for the gamma function}
\end{equation}

To be explicit we calculate the $A=T_{+}^{+}$ element by using its
logarithm. Then the logarithm of infinite products turns into infinite
summations, and we can use the $\left( \ref{integral representation for the
gamma function}\right) $ formula.

After some calculation we reach a formula, which can be rearranged in terms
of trigonometric and hyperbolic functions%
\begin{equation}
A(\theta )=\frac{1}{\sqrt{\nu }}\frac{e^{i\pi (1+\gamma )/4}}{(1+ipx)}\exp
\left( i\int_{0}^{\infty }\frac{dt}{t}\frac{\sin \left( \frac{\gamma \tilde{%
\theta}t}{\pi }\right) \sinh \left( \frac{t}{2}(\gamma +1)\right) }{\sinh
\left( t\right) \cosh \left( \frac{\gamma t}{2}\right) }\right)
\label{integral representation for T even A element}
\end{equation}

If we want to separate the $\varphi $ phase shift we have to concentrate on
the argument of the exponential in the form $e^{i\varphi }$. The
semiclassical approximation means that in the integrand we substitute the
component functions with its Taylor-series. Then we keep only the leading
order in the $\gamma $ coupling constant.

After making this process the resulting expression for the phase shift%
\begin{equation}
\varphi =\int_{0}^{\infty }\frac{dt}{t^{2}}\sin \left( (\theta -\eta
)t\right) \tanh \left( \frac{1}{2}\pi t\right) \frac{\gamma }{\pi }
\label{semiclassical limit off T even}
\end{equation}

We can use the $\left( \ref{coupling constants}\right) $ connection between
the coupling constants in order to express $\gamma $ with $\beta $ in the
leading order%
\begin{equation}
\frac{1}{\gamma }=\frac{\beta ^{2}}{8\pi }\left( \frac{1}{1-8\pi /\beta ^{2}}%
\right) =\frac{\beta ^{2}}{8\pi }\left( 1+\frac{8\pi }{\beta ^{2}}+\left( 
\frac{8\pi }{\beta ^{2}}\right) ^{2}+...\right)
\end{equation}

In the leading order it follows that%
\begin{equation*}
\gamma =\frac{8\pi }{\beta ^{2}}
\end{equation*}

Then we can express $\left( \ref{semiclassical limit off T even}\right) $

\begin{equation}
\varphi =\frac{8}{\beta ^{2}}\int_{0}^{\infty }\frac{dt}{t^{2}}\sin \left(
(\theta -\eta )t\right) \tanh \left( \frac{1}{2}\pi t\right) \mathfrak{+%
\mathcal{O}(\beta ^{0})}  \label{semiclassical limit off T even (other form)}
\end{equation}

\subsection{Comparison of the two results}

The final equation in question come from $\left( \ref{phase shift on the
even defect}\right) $ and $\left( \ref{semiclassical limit off T even}%
\right) $: 
\begin{equation}
\dint\limits_{0}^{\infty }\frac{dt}{t^{2}}\sin \left( (\theta -\eta
)t\right) \tanh \left( \frac{1}{2}\pi t\right) =\dint\limits_{0}^{\theta
}d\theta ^{\prime }\ln \left( \tanh \frac{(\eta -\theta ^{\prime })}{2}%
\right)  \label{the final equation}
\end{equation}

Differentiating both sides with respect to $\theta $:%
\begin{equation}
\dint\limits_{0}^{\infty }\frac{dt}{t}\cos \left( (\theta -\eta )t\right)
\tanh \left( \frac{1}{2}\pi t\right) =\ln \left( \frac{\eta -\theta }{2}%
\right)  \label{differentiating the final equation}
\end{equation}

This is a known identity, which can be found in table of integrals, for
example in \cite{key-19}. We have proven that the two semiclassical limit
give the \textit{same }result. So we can assume that the unknown CDD scalar
function is the simplest one.

\subsection{The case of odd charged defects}

If the defect had an odd charge we have to start the calculation from the $%
\left( \ref{common function in the odd case}\right) $ expression. Using our $%
\left( \ref{integral representation for the gamma function}\right) $
integral representation for the gammas the result for the exponential in $%
\hat{A}(\theta )$%
\begin{equation}
\exp \left( i\int_{0}^{\infty }\frac{dt}{t}\frac{-\sin \left( \gamma \frac{%
\tilde{\theta t}}{\pi }\right) \sinh \left( \frac{1}{2}t(\gamma -1)\right) }{%
\sinh \left( t\right) \cosh \left( \frac{t\gamma }{2}\right) }\right)
\label{integral representation for T odd A element}
\end{equation}%
\qquad \qquad

Let us compare this formula with the earlier and very similar $\left( \ref%
{integral representation for T even A element}\right) $ expression which we
found in the case of the even charged defect. The difference in the argument
of the hyperbolic function - $+$ or $-$ sign - doesn't count in the $\gamma
\rightarrow \infty $ semiclassical limit. So the new formula differs only a
minus sign compared to $\left( \ref{semiclassical limit off T even (other
form)}\right) $%
\begin{equation}
\varphi =-\frac{8}{\beta ^{2}}\int_{0}^{\infty }\frac{dt}{t^{2}}\sin \left(
(\theta -\eta )t\right) \tanh \left( \frac{1}{2}\pi t\right) \mathfrak{+%
\mathcal{O}(\beta ^{0})}
\end{equation}

We have found this minus sign in the $\left( \ref{time delay on the odd type
defect}\right) $ time delay of the odd charged defect. In this way we can
trace back the checking of the phase shifts\ to the even case: we found that
the two result is \textit{equal} again.

\bigskip

After this we examine the structure of the even transmission matrix $^{e}T$
in the semiclassical limit, which was given by $\left( \ref{T matrix even}%
\right) $. In this limit the quantum features of the model disappear and
graduate into classical behavior. If we want unambiguous result we must fix
the relative value of the parameters of motion: the rapidity of the soliton
and the parallel defect parameter.

Assume that%
\begin{equation}
\theta >\eta \Longleftrightarrow \theta -\eta >0
\label{relation between paramters : case 1}
\end{equation}

and suppose that the $\gamma $ coupling constant is going to infinity. In
this limit we find that the limit value of the off-diagonal element of the
transmission matrix 
\begin{equation}
\frac{e^{i\pi (1+\gamma )/4}e^{-i\pi \gamma /2+\gamma (\theta -\eta )}}{%
(1+ie^{\gamma (\theta -\eta )})}=-\frac{e^{-i\pi \gamma /4}e^{\gamma (\theta
-\eta )}}{(1+ie^{\gamma (\theta -\eta )})}\rightarrow ie^{-i\pi \gamma /4}
\label{limit value of the off-diagonal : 1. case}
\end{equation}

It is important to note that the exponent is proportional to the coupling
constant, which can be an indication that bound states or resonances exists
in the spectrum. In the classical case using $\left( \ref{time delay with
the presence of a defect}\right) $ we saw that there is no upper bound of
the time delay. But it doesn't mean necessarily that there are bound states
in the quantum version. If we consider the phenomenon of quantum tunnelling
maybe we can find only short life resonances.

In the same limit the diagonal elements of the matrix goes to zero:%
\begin{equation}
\left\vert \frac{e^{i\pi (1+\gamma )/4}}{(1+ie^{\gamma (\theta -\eta )})}%
\right\vert =\frac{1}{\left\vert 1+ie^{\gamma (\theta -\eta )}\right\vert }%
\leq \frac{1}{\left\vert e^{\gamma (\theta -\eta )}\right\vert }\rightarrow
0,\quad \gamma \rightarrow \infty
\label{limit value of the diagonal : 1. case}
\end{equation}

So the matrix has the following symbolic structure in the semiclassical limit%
\begin{equation}
\left( 
\begin{array}{cc}
0 & \bullet \\ 
\bullet & 0%
\end{array}%
\right)  \label{matrix structure semiclassical : case 1}
\end{equation}%
which shows that in this case the type of the soliton always changes, when
it goes over the defect.

Reverse the relation between the parameters, that is to say 
\begin{equation}
\theta <\eta \Longleftrightarrow \theta -\eta <0
\label{relation between paramters : case 2}
\end{equation}

In this case the situation changes, the off-diagonal elements go to zero%
\begin{equation}
\left\vert \frac{e^{i\pi (1-\gamma )/4}\overset{\rightarrow 0}{\overbrace{%
e^{\gamma (\theta -\eta )}}}}{(1+i\underset{\rightarrow 0}{\underbrace{%
e^{\gamma (\theta -\eta )}}})}\right\vert \rightarrow \left\vert \frac{%
e^{i\pi (1-\gamma )/4}0}{(1+0)}\right\vert =0
\label{limit value of the off-diagonal : 2. case}
\end{equation}

and the diagonal elements have the finite value:%
\begin{equation}
\frac{e^{i\pi (1+\gamma )/4}}{(1+ie^{\gamma (\theta -\eta )})}=-e^{i\pi
\gamma /4}  \label{limit value of the diagonal : 2. case}
\end{equation}

The structure we found for the transmission matrix%
\begin{equation}
\left( 
\begin{array}{cc}
\bullet & 0 \\ 
0 & \bullet%
\end{array}%
\right)  \label{matrix structure semiclassical : case 2}
\end{equation}
which represents that the soliton type remain unchanged.

\section{Resonances}

The analysis of the structure of the even transmission matrix $^{e}T$ shows
that bound states or resonances can appear in the spectrum of the
transmission. The limit values in the semiclassical limit were proportional
to the coupling.

This phenomena can be examined directly, by extracting the pole structure of
the transmission matrix. To do this we invoke that a gamma function has a
simple pole at every non positive integer. But we have to take care because
there are poles in denominator of the $\left( \ref{T matrix even}\right) $
formula: if the pole of the denominator is at the same location as the pole
of the numerator, then the pole in question has been cancelled out.

Firstly we would try to find the bound states. This means that we have to
locate the poles in the so-called "physical strip" \cite{bs5}. It means that
the rapidity should be in the $[0,i\pi /2)$ domain of the complex plane.

From the $\left( \ref{common function}\right) $ and the expression $\left( %
\ref{gammas product in the even T matrix}\right) $ it is clear that there
are four \textit{indexed} gamma function in the numerator, and four similar
in the denominator. The eight equations for the poles 
\begin{eqnarray}
k\gamma +\frac{1}{4}\mp \frac{i\gamma \widetilde{\theta }}{2\pi } &=&-n
\label{the poles of the numerator in even} \\
(k+1)\gamma +\frac{3}{4}\mp \frac{i\gamma \widetilde{\theta }}{2\pi } &=&-n 
\notag \\
(k+\frac{1}{2})\gamma +\frac{1}{4}\pm \frac{i\gamma \widetilde{\theta }}{%
2\pi } &=&-n  \notag \\
(k+\frac{1}{2})\gamma +\frac{3}{4}\pm \frac{i\gamma \widetilde{\theta }}{%
2\pi } &=&-n  \notag
\end{eqnarray}%
where $n=0,1,2,...$. The upper sign denotes the numerator, and the lower the
denominator.

After expressing the rapidity we found that the $n$-th pole of the $k$-th
gamma is at rapidity

\begin{eqnarray}
\widetilde{\theta } &=&\mp i\left( 2\pi k+\frac{\pi }{2\gamma }(4n+1)\right)
\label{the location of the poles} \\
\widetilde{\theta } &=&\mp i\left( 2\pi (k+1)+\frac{\pi }{2\gamma }%
(3+4n)\right)  \notag \\
\widetilde{\theta } &=&\pm i\left( \pi (2k+1)+\frac{\pi }{2\gamma }\left(
4n+1\right) \right)  \notag \\
\widetilde{\theta } &=&\pm i\left( \pi (2k+1)+\frac{\pi }{2\gamma }%
(4n+3)\right)  \notag
\end{eqnarray}%
where the notation of the signs is the same. ($\widetilde{\theta }=\theta
-\eta $ but $\eta $ is real, so it doesn't effect the domain). Taking into
account the poles of the denominator we found that there are no poles in the
"physical strip".

After this we turn to the examination of so-called "resonance" poles \cite%
{bs5}. It means that we should locate poles with rapidity in the $[0,-i\pi
/2)$ interval. We can use the $\left( \ref{the location of the poles}\right) 
$ formulas without change, together with the similar expressions for the
denominator.

In this case we found poles in the given domain. In the semiclassical limit $%
\gamma \rightarrow \infty $ the number of poles is proportional to the
coupling $\gamma $%
\begin{equation}
n_{p}\sim \frac{\gamma }{4}
\end{equation}

This relation agree with the expressions $\left( \ref{limit value of the
off-diagonal : 1. case}\right) $ and $\left( \ref{limit value of the
diagonal : 2. case}\right) $ found in the semiclassical limit of the even
transmission matrix.

\section{Conclusion}

In this article an interesting version of the nonlinear sine-Gordon field
theory has been examined. The theory includes an integrable jump-defect in
the origin, which means that the integrability - an important characteristic
of the original model - is not destroyed.

It is required that the T-matrix of the defect theory satisfy the
Yang-Baxter equation (also known as "factorisability condition"), and the
standard equation of unitarity and crossing symmetry. These equation have
enough restrictive power to determine the T-matrix up to the so-called "CDD
ambiguity".

It means that there is an unknown scalar function in the scattering and the
defect transmission matrix too. This function cannot be fixed via the
bootstrap method. One way to fix this ambiguity is to calculate the phase
shift by the semiclassical approximation. In the case of quantum sine-Gordon
theory the classical phase shift has been compared with the scattering
matrix earlier \cite{key-15}, \cite{key-24}. One of the aims of this article
is to determine the phase shift in the case of \textit{defect} sine-Gordon
theory. Using the result we can anchor the CDD uncertainty. The defect
theory has new symmetries so we had to recalculate the time delay for a new
case.

Another objective is to map the pole structure of the T-matrix. In this way
we can study the spectrum of the theory if there are stable or unstable
bounds states or resonances.

In the above sections we have made the semiclassical limit, and we have
found that in the leading order of the semiclassical approximation the
classical and the quantum field theory agree: the phase shifts are equal.
The unknown scalar function can be the trivial one.

The classical results confirmed that a defect - effectively - behaves like a
half soliton: the time delay is half of time delay if the soliton scatters
on an antisoliton. Naturally the result for the phase shift is the same. The
similarity in the quantum version remained an open question.

We examine the structure of the transmission matrix in the semiclassical
limit too. The results indicate that there can be stable or unstable bound
states or resonances in the spectrum. The remaining matrix entries are the
function of the quarter of the coupling constant, which shows the
half-soliton behavior again.

But the direct inspection demonstrated that there are no bound states in the
spectrum of a sine-Gordon theory with a jump-defect. But there are
resonances in the model and naturally the number of resonances is
proportional to the coupling constant.

\bigskip \textbf{Acknowledgements}

\smallskip I would like to thank Zolt\'{a}n Bajnok who help during all of my
work with lots of professional advices.

\end{document}